\documentclass[twocolumn,aps,pre,epsfig,amsmath,eqsecnum]{revtex4}
\usepackage{dcolumn}
\usepackage[dvips]{graphicx}

\date{\today}

\begin{document}
\title{Statistical mechanics of RNA folding:
importance of alphabet size}

\author{Ranjan Mukhopadhyay, Eldon Emberly\footnote{Current 
Address: Center for Physics and Biology, Rockefeller University, New York,
NY 10028}, Chao Tang, and Ned S. Wingreen}
\affiliation{NEC Laboratories America, Inc., 4 Independence Way, Princeton, NJ 08540}
\date{\today}

\begin{abstract}
 We construct a base-stacking model of RNA secondary-structure formation
and use it to study the mapping from sequence to structure. 
There are strong, qualitative differences between two-letter and four or
six-letter
alphabets. With only two kinds of bases, most sequences have many alternative 
folding configurations and are consequently thermally unstable.
Stable ground states are found
only for a small set of structures of high designability, {\it i.e.}
total number of associated sequences. In contrast, sequences
made from four bases, as found in nature, or six bases have far 
fewer competing folding
configurations, resulting in a much greater average stability of the 
ground state.
\end{abstract}


\maketitle


\section{Introduction}

RNA plays a central role in molecular biology.
In addition to transmitting genetic information from DNA to  
proteins, RNA molecules participate actively in a variety of cellular
processes~\cite{RNAworld}. Examples are found in translation (rRNA, 
tRNA, and tmRNA),  
editing of mRNA, intracellular protein targeting, nuclear splicing of  
pre-mRNA, and X-chromosome inactivation.
The RNA molecules involved in these processes do not code for proteins but
act as  functional products in their own right. In addition,  
RNA molecules prepared {\it in vitro}
can be selected to bind to specific molecules such as ATP~\cite{RNAaptamers}.  
In all these cases, the information encoded in the sequence 
of nucleotide bases of each RNA molecule determines its
functional three-dimensional structure. The nucleotide sequence
is a kind of genotype, {\it i.e.}, hereditary information, while
the folded three-dimensional structure represents phenotype, the
physical characteristics on which natural selection
operates. The mapping from genotype to phenotype bears
on how biological systems evolve, and 
RNA folding probably constitutes the simplest example of this 
mapping~\cite{fontana1}.
 Since early life is believed to
have been RNA based~\cite{RNAworld}, RNA folding can provide us with 
important  clues about early life and evolution.

   RNA is a polynucleotide chain consisting of the four bases:
A, U, G, and C. Complementary  base pairs (A-U and G-C) can stack to
form ``stems'' which are helical segments similar to the double helix of DNA. 
These helices, called secondary structures, are generally arranged 
in a three-dimensional tertiary  
structure, stabilized by the much weaker interactions between the helices.
Representations of secondary structures are shown in Fig.~1. 
The energy contributions of secondary and tertiary structures are
hierarchical\cite{bustamante}, with secondary structures largely determining 
tertiary folding. Secondary structure is frequently conserved in
evolution, and structural homology has been used successfully to predict 
function \cite{zuker1}.

In this paper, we investigate the role of alphabet size in 
the statistical mechanics and selection of RNA secondary structures.
We find pronounced differences between two-letter and four
or six-letter
alphabets. For sequences constructed with two types of bases, 
only a small fraction of sequences have thermodynamically
stable ground-state structures; these structures are also
highly designable, {\it i.e.}, have a large number of associated sequences. 
Four and six-letter sequences are much more stable on    
average, but exhibit
no strong correlation between designability and thermodynamic
stability. We trace this difference to the greater likelihood
of competing, alternatively paired configurations when a two-letter 
alphabet is used.

For RNA, there already exist algorithms that predict secondary 
structures \cite{zuker,viennapackage}.
These algorithms are intended to apply to real RNA and,
consequently, involve a large number of parameters for the
different pairing and stacking combinations. Using
one of these algorithms, Fontana {\it et al.}\cite{fontana}
found a broad distribution of designabilities, {\it i.e.}
number of sequences per structure, after structures
were grouped by topology. In this paper, we present,
instead, a much simpler
model for RNA secondary structure designed to elucidate the 
role of alphabet size.

  The organization of this paper is as follows. In section II, we present 
a base-stacking model for RNA secondary structure and outline 
the recursive algorithm used to compute the partition function and 
ground-state structure. In section III, we employ our model to analyze 
the stability of folded structures. We find a significant difference
in stability between two-letter and four or six-letter sequences due to the greater 
likelihood of alternative folds in the two-letter case. As a consequence
of these alternative folds, in the two-letter case, stability correlates
with designability, {\it i.e.}, total number of sequences associated with a structure.
In addition, we find that RNA sequences folding to a given structure form a percolating neutral
network. Finally, in section IV, we summarize our main conclusions.

\section{Base-stacking Model}

\begin{figure}[!hbp]
\begin{center}
\includegraphics[width=7cm]{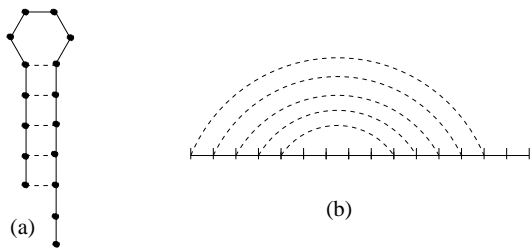}
\vspace{0.25cm}
\caption{ 
Representations of RNA secondary structures: (a) flattened-helix
diagram of a 16-base structure, and (b) rainbow diagram
of the same structure. The restriction that arches do not cross in the 
rainbow diagram implies the absence of pseudoknots.}
\label{fig:Fig1}
\end{center}
\end{figure}

We introduce a base-stacking model for RNA secondary-structure  
formation. It is known that, within a stem
of base pairs, the largest energy contribution is the {\it stacking  
energy} between two adjacent base pairs (rather than the base-pairing
energy itself) and the total energy of the stem is the sum of stacking  
energies over all adjacent base pairs~\cite{bustamante}. A single stack
($i,i+1;j-1,j$) is defined as two adjacent non-overlapping base pairs 
($i,j$) and ($i+1,j-1$) where $i+1<j-1$.
For this stack ($i,i+1;j-1,j$)
we assign an energy $-E_{s}$ if ($i,j$) and ($i+1,j-1$) are
both complementary Watson-Crick base pairs and zero otherwise.  
We thus neglect differences in energy between, for example,
(A,A;U,U), (A,G;C,U) and (G,G;C,C) stacks. We also neglect energy
contributions from isolated base pairs that are not part of a stack,
and, consequently, do not include 
isolated base pairs in the secondary structure.

The largest entropic contribution to an RNA structure comes from
stretches of unpaired bases. We incorporate a simplified version of 
this polymer configurational entropy in our model by
associating $\alpha$ degrees of freedom with every unpaired base. Thus,
the restricted partition function, corresponding to all micro-states compatible
with a given secondary structure is
\begin{equation}
Z_{\rm micro} = \alpha^{n_{u}}  
\exp \left[{\frac {n_{s} E_{s}} {k_{B} T}}\right] 
\end{equation}
where $n_{u}$  
is the number of unpaired bases, $n_{s}$ is the number of stacks, and $T$ is
the temperature. The restricted free  
energy is $F_{\rm micro}= -k_BT \ln Z_{\rm micro} = 
- E_{s}n_{s} - k_{B} T n_{u} \ln \alpha$. 

In this model, since
only complementary base pairs can participate in a stack,
only a fraction of possible structures are compatible with any given
sequence. However, provided the structure is compatible with the
sequence, its restricted free
energy is independent of the sequence.

The change in free energy due to the formation of an isolated stack is  
$-E_{s} + 4  k_{B} T \ln \alpha$; the first term corresponds 
to the stacking energy and the second to the loss in configurational  
entropy (since four bases participate in the stack).  
For every additional adjacent stack the change in free energy is
$- E_{s} + 2 k_{B} T \ln \alpha$, since only two bases are
added to the stack. If, for example, $E_{s} < 4 k_B T\ln \alpha$
but $2 E_s > 6 k_B T \ln \alpha$ ({\it i.e.}, $ 3k_BT \ln \alpha <
E_s < 4k_BT \ln \alpha)$,
then formation of an isolated stack
would be unfavorable but formation of a segment consisting of two or more  
adjacent stacks would be favored by a net decrease in free energy. 
Thus, for an appropriate
choice of parameters, the model correctly provides a  
nucleation cost to the formation of stems. For this paper we choose
$\ln \alpha=1.5$ and $E_{s} = 5.5 k_B T$, which are physically motivated
and correspond to a nucleation cost for the formation of an isolated stack,
with a minimum of two adjacent stacks required to form a stable stem.
Our results, however, do not depend sensitively on the choice of these 
parameters. 


In the secondary structure, any two base pairs 
($i_{1},j_{1}$)  and ($i_{2},j_{2}$), with $i_{1}<i_{2}$, are either nested  
($i_{1} < i_{2} < j_{2} <j_{1}$) or independent
($i_{1} < j_{1} < i_{2} <j_{2}$). Other possibilities 
correspond to ``pseudoknots'', which are energetically
and kinetically suppressed. It is customary to regard pseudoknots as 
part of the tertiary structure, and we do not include them here.

       In order to compute the ground-state structure and
partition function for a given sequence, we make use of the hierarchical
nature of secondary structures (due to the absence of pseudoknots).
We use a recursive algorithm
that is a generalization of the techniques described in Refs. 
\cite{recursion1} and \cite{hwa}.
Consider the partition function $Z_{i,j}$ for
a segment of bases from the position $i$ to $j \ge i$. The base $j$
is either unpaired or can be part of a stack $(k,k+1;j-1,j)$
with  $k \in \{i, \ldots , j-3\}$. Thus $Z_{i,j}$ obeys:
\begin{eqnarray}
Z_{i,j} &=& \alpha Z_{i,j-1} + \sum_{k=i}^{j-3} [Z_{i,k-1}
\cdot e^{E_{s}/k_{B}T} \nonumber \\
& & \cdot {\mathcal P}_{s}(k,k+1;j-1,j) \cdot {\hat Z}_{k+2;j-2}],
\end{eqnarray}
where ${\mathcal P}_{s}(k,k+1;j-1,j)$ equals~1 if both $(k,j)$ 
and $(k+1,j-1)$ are
complementary base pairs, and equals~0 otherwise; $Z_{i,i-1}$ is defined
to equal $1$. We have introduced
${\hat Z}_{i,j}$ which is the partition function for the segment
with the boundary condition that sites $i-1$ and $j+1$ are paired,
implying an energy $-E_{s}$ for the formation of a bond between
the bases at sites $i$ and $j$. We thus require a second recursion
relation for ${\hat Z}$:
\begin{eqnarray}
{\hat Z}_{i,j} &=& \alpha Z_{i,j-1} + e^{E_{s}/k_{B}T} \cdot 
{\mathcal P}_{b}(i,j) \cdot 
{\hat Z}_{i+1,j-1}  \\ \label{zij}
&+&  \sum_{k=i+1}^{j-3}[ Z_{i,k-1}
e^{E_{s}/k_{B}T} {\mathcal P}_{s}(k,k+1;j-1,j)
\cdot {\hat Z}_{k+2;j-2}], \nonumber
\end{eqnarray}
where ${\mathcal P}_{b}(i,j)$ equals 1 if $(i,j)$ are complementary base
pairs and $0$ otherwise. The partition function $Z_{1,N}$ can
be computed recursively using (2) and (3) in $O(N^{3})$ steps. We use
a similar recursive algorithm to compute the ground-state structure 
${\mathcal S}_{1,N}$.

\begin{figure}[!hbp]
\begin{center}
\includegraphics[width=7cm]{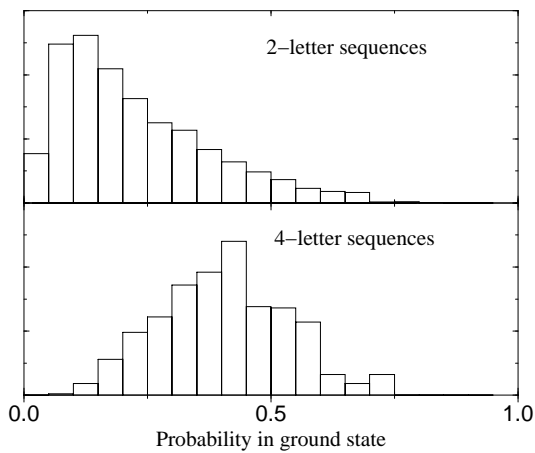}
\vspace{0.25cm}
\caption{Histogram of stabilities for: 
(a) 40-nucleotide long two-letter sequences, and
(b) 40-nucleotide long four-letter sequences.}
\label{fig:Fig2}
\end{center}
\end{figure}

\section{Results}

\subsection{Dependence on Alphabet Size}

  We have employed our model to analyze the stability of folded
structures corresponding to two, four, and six-letter sequences. 
The thermodynamic stability is defined as the
probability $P_{\rm GS}$ that the sequence will be found in the ground state,
$P_{\rm GS} = e^{- F_{\rm GS}/k_{B} T}/ Z$ where  
$F_{\rm GS}$ is the free energy associated with the ground state. 
Fig.~2 shows a histogram
of stability  for 40-nucleotide long sequences with
ground states containing 12 to 15 stacks~\cite{footnote1}. We find four-letter
sequences considerably more stable on average than two-letter sequences.

\begin{figure}[!hbp]
\begin{center}
\includegraphics[width=7cm]{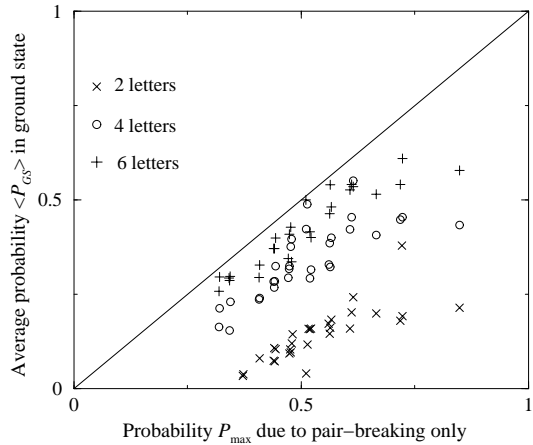}
\vspace{0.25cm}
\caption{Average of actual probability in ground state vs. upper bound 
$P_{\rm max}$ allowing only for breaking of base pairs, plotted
for 40-nucleotide sequences.
$+$'s denote RNA sequences constructed from two types of bases,
$\circ$'s denote those constructed from four, and
$\times$'s denote sequences constructed from six types of bases.
The actual probability is averaged over sequences 
with the same pair-breaking stability  
$P_{\rm max}$ (which is sequence independent).}
\label{fig:Fig3}
\end{center}
\end{figure}

   What is the origin of the difference in stabilities between  
two-letter and four-letter RNA sequences? 
In order to address this question, we classify 
the excited-state structures as (i) those formed
by breaking existing pairs, and (ii) those formed by re-pairing, {\it i.e.}, by
forming new pairs in addition to breaking existing pairs. 
Independent of alphabet size,
all sequences folding into a given secondary structure ${\mathcal S}$ 
have the same set of ``pair-breaking'' excited states. 
The {\it sequence} dependence of stability for a given 
ground-state structure results entirely from re-pairings. 
{\it The crucial difference between two-letter and four 
sequences lies in the substantially greater likelihood of ``re-paired'' 
excited states for two-letter sequences.} 
This follows because the number of pairs one can form  
in a random sequence of two letters is typically much larger than
for a four or six-letter sequence of the same length. For example,
for a random four-letter sequence of length $N$, 
the probability of forming a stem
involving sites $i$ to $i+l$ and $j-l$ to $j$
is lower by a factor of $2^{l}$ 
as compared to a random two-letter sequence of the same length.
For the same reason, the fraction of sequences that have highly
stacked ground states
is much greater for two-letter sequences than for four-letter
sequences, and much greater for four than for six.

  To demonstrate the importance of ``re-paired' excited states,
we first calculate a ``pair-breaking'' stability 
$P_{\rm max} = e^{-F_{\rm GS}/k_{B} T}/{\mathcal Z}$ where
${\mathcal Z}$ is a pair-breaking partition function calculated
by considering only pair-broken excited states. $P_{\rm max}$
gives us an upper bound to the true stability, {\it i.e.}, 
probability in ground state $P_{\rm GS}$, that includes competition
from re-paired states.
In Fig.~3, we plot the true average stability $\langle P_{\rm GS}\rangle$ 
against the pair-breaking stability  
$P_{\rm max}$ for two, four, and six-letter sequences. 
As expected, the average stability is much closer to the maximum
set by pair breaking in the case of four-letter sequences than in
the case of two-letter sequences.  Thus, structures constructed with  
four-letter sequences are typically much more than stable
than those constructed with two letters, and six-letter sequences 
are typically more stable than four-letter ones.
For folding {\it kinetics},
it is these same ``re-paired''  states that act as kinetic traps.
Due to the lower likelihood of such states, we expect four and six-letter
sequences to typically fold faster than two-letter sequences.

\subsection{Stability and Designability} 
  What determines the average stability, $\langle P_{GS}\rangle$, 
of two-letter sequences? 
We have seen that for four and six-letter sequences the average
stability is close to the ``pair-breaking'' stability which
is determined largely by the number of stems and loops. 
Insight into the stability of two-letter RNA sequences
comes from  results in protein folding
\cite{Chan,lhtw,ltw-pnas,regis,shak99,ltw2}. 
Based on solvation models with
differing hydrophobicities of amino acids, a principle of 
designability has emerged for protein folding.
The designability of a structure is measured 
by the number of sequences folding uniquely into that structure.
A small class of protein stuctures emerge as being highly designable;
remarkably, the same class of structures are highly designable whether two or 
all 20 amino-acid types are used\cite{ltw2}.
In a wide range of protein models, sequences associated with
highly designable structures are thermodynamically more 
stable\cite{lhtw,mzwt} and fold faster than typical sequences\cite{regis}. 
This connection between 
the designability of a structure and the stability of its associated
sequences is referred to as the designability principle. 
The designability principle reflects a competition among structures.
In solvation models, sequences will fold to structures
which best match their hydrophobic amino acids to buried
sites in the structure (shielded from water). Highly designable 
structures are those with unusual patterns of surface
exposure, and therefore few competitors. This lack of competitors also
implies that the sequences folding to such structures are thermally stable. 
We will now show that the designability principle 
also holds for two-letter RNA.

\begin{figure}[!hbp]
\begin{center}
\includegraphics[width=7cm]{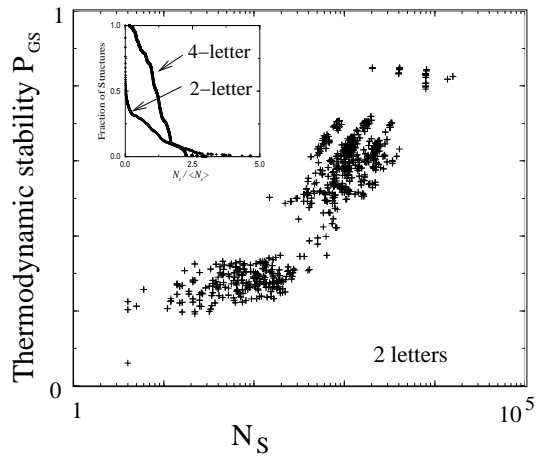}
\vspace{0.25cm}
\caption{Stability [20] versus designability $N_{s}$ (in logarithmic
scale) for 24-base RNA sequences constructed
with two types of bases. In the inset we plot fraction of compact 
structures [19] with designability above $N_{s}$ versus $N_{s}$
for two and four-letter RNA sequences.}
\label{fig:Fig4}
\end{center}
\end{figure}

\begin{figure}[!hbp]
\begin{center}
\includegraphics[width=7cm]{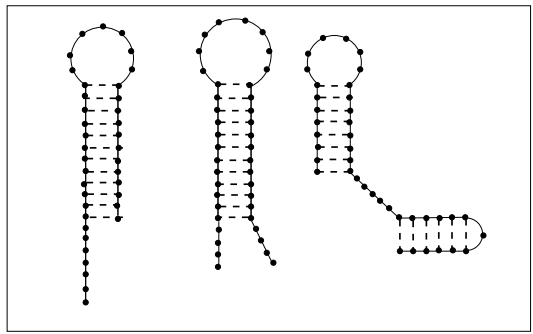}
\vspace{0.25cm}
\caption{A few highly designable structures for 40-nucleotide
long two-letter RNA sequences.}
\label{fig:Fig5}
\end{center}
\end{figure}

For two base-types (say A and U), we 
enumerate all sequences and structures of length 24.
We find that
secondary structures differ considerably in their designability; there
are highly designable structures which are ground states of a large 
number of sequences, and there are poorly designable structures which
are ground states of only a few sequences (cf. Fig.~4 inset).  
In this respect, the results
for two-letter sequences are similar to those for protein 
models\cite{lhtw,ltw-pnas}. However, the histogram is
more noisy for RNA than it is for proteins; so we plot the
integrated distribution of designabilities. The most designable
structure consists of a stem with a hairpin loop, and a dangling end.
We have also studied longer sequences, of lengths 40 and 50,
for which we sample sequence space. For 40-nucleotide sequences,
the most designable structures  
consist of a single hairpin loop and dangling ends; a number of
double hairpin structures are also highly designable (Fig. 5). 
For sequences of length 50,
double hairpin structures emerge as the most designable.
Finally, we find a pronounced correlation between designability and 
stability of RNA structures. This is
shown in Fig.~4~\cite{footnote2} for 24-nucleotide sequences. 
Thus, two-lette RNA sequences which fold 
into highly designable secondary structures are unusually
thermally stable, verifying the designability principle. 

\begin{figure}[!hbp]
\begin{center}
\includegraphics[width=7cm]{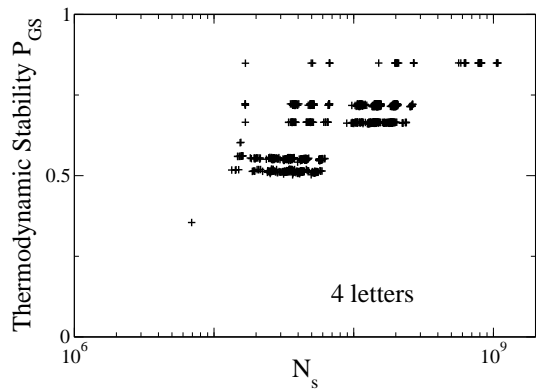}
\vspace{0.25cm}
\caption{Stability [20] versus designability $N_{s}$ (in logarithmic
scale) for 24-base RNA sequences constructed
with four types of bases. We find no significant correlation between
designibality and stability for four letters. }
\label{fig:Fig6}
\end{center}
\end{figure}

In contrast, for four-letter sequences
the range of designabilities is
narrower and there is only
a weak correlation between designability and stability, with highly
stable sequences existing for structures of both high and low
designability (Fig.~6). The results for six letters are similar.
We trace this difference between two and four
or six-letter
sequences to the likelihood of competing re-paired states. 
For two letters, the correlation between designability and stability 
(as well as the nontrivial distribution of designabilities) arises
primarily from competing re-paired states. Four and six-letter
sequences have far fewer competing re-paired states and 
hence do not demonstrate
significant correlation between designability and stability.

\subsection{Neutral Networks}

\begin{figure}
\begin{center}
\includegraphics[width=7cm]{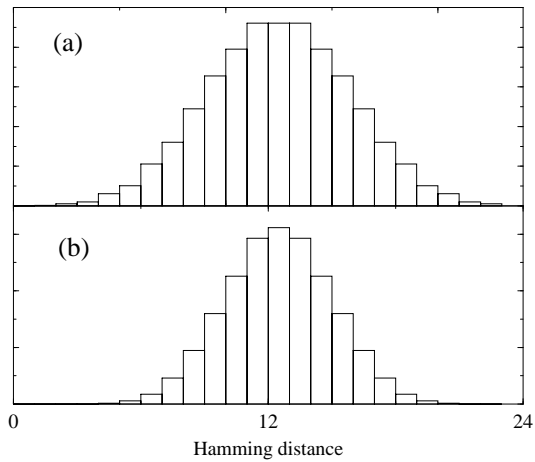}
\vspace{0.25cm}
\caption{For some given 24-nucleotide two letter sequence $\sigma_{1}$ we plot 
(a) a histogram of the
distances to all two-letter sequences with the same ground state structure, and
(b) a histogram of the distances to all two-letter sequences.
Histogram (b) is independent of the choice of $\sigma_{1}$.
Histogram (a) is also roughly independent
of sequence $\sigma_{1}$ provided its ground-state structure 
is highly designable.}
\label{fig:Fig7}
\end{center}
\end{figure}

Finally we consider the ``neutral network'' of RNA sequences
which fold to a particular structure. The connectivity within
a network and the shortest distance between networks 
has drawn considerable attention with respect to the 
evolvability of RNA structures\cite{fontana,fontana2}.
In our model, the network of sequences which fold to a particular
structure is truly ``neutral'' in that all sequences 
have the same ground-state free energy $F_{\rm GS}$, albeit
with different stabilities because of repairing.
(This contrasts with protein solvation models in which, independent
of competing structures, there
is typically an energy hierarchy of sequences for each structure,
determined by the match between hydrophobicity and surface-exposure 
pattern\cite{ltw-pnas}.) In our model, RNA sequences that
fold to a given structure form, in general, a percolating
and non-compact network in sequence space. In particular, a 
histogram of the distances 
between sequences folding to the
same highly designable structure is actually broader than
a histogram of the distances between {\it all} sequences
(Fig.~7)\cite{hamming}.
In this respect, the RNA model differs considerably from protein
models. 
 
\section{Conclusions}

       To conclude, in this paper we developed and studied
a minimalist base-stacking model of RNA secondary structure. 
We found that sequences
constructed with four or six types of bases typically have fewer 
competing excited states, and, consequently, have greater 
ground-state stability, compared to sequences constructed with two types
-of bases. At the same time, the fraction of sequences with highly
stacked ground states is much smaller for four-letter sequences
than for two, and  much smaller for six letters than for four. 
It is tempting to speculate that four letters 
optimizes the stability of structures while maintaining a   
reasonable probability that a random sequence folds into a highly
stacked structure.  If, as has been postulated, early life was indeed RNA
based and double-stranded DNA came later in evolution,
our observations might plausibly
bear on nature's choice of four letters for the genetic code.

\begin{acknowledgments}
We thank David Moroz for useful discussions and suggestions.
\end{acknowledgments}



\begin{thebibliography}{1}
\bibitem{footnote1}
We use a narrow range of stack numbers to emphasize the dependence
of stability on alphabet size. For a wider range of stack numbers,
the increase in stability with stack numbers can obscure
the dependence on alphabet sizes.
\bibitem{sample}
For each structure,
we generate a random sample of sequences that are compatible with the structure
and calculate the fraction of such sequences that have this structure as 
the ground state. We multiply the total number of compatible sequences
by this fraction to obtain the designability.
\bibitem{compact} In Fig.~4, we have kept only structures that are
highly stacked, in particular, those that have six or fewer unpaired bases.
Our results do not depend sensitively  on this cutoff.
\bibitem{footnote2}
In Fig.~4, we plot the
$90$-th percentile of greatest stability, $P_{90\%}$, rather than average 
stability of sequences folding to a structure. Since sequences
folding to a structure have, in general, a wide range of 
stabilities, the two can be quite different. Average
stability shows a similar, but less pronounced correlation with designability
than $P_{90\%}$.
\bibitem{hamming} A ``Hamming'' distance between RNA sequences can be
defined as a distance of 1 between non-identical bases at a site and 
0 between identical bases.
\end{thebibliography}
\end{document}